\documentclass[11pt,a4paper]{article}

\usepackage{amsmath}
\usepackage{amsthm,amssymb}

\usepackage{graphicx}
\usepackage{grffile}

\usepackage{tikz}
\usetikzlibrary{decorations}
\usetikzlibrary{decorations.pathmorphing}
\usetikzlibrary{scopes}

\newcommand{\beq}{\begin{equation}}
\newcommand{\eeq}{\end{equation}}

\newcommand{\<}{\langle}
\renewcommand{\>}{\rangle}

\DeclareMathOperator*{\argmax}{arg\,max}

\numberwithin{equation}{section}

\begin{document}
\title{Self-avoiding trails with nearest neighbour interactions on the square lattice}
\author{A Bedini$^1$, A L Owczarek$^1$ and T Prellberg$^2$\\
  \footnotesize
  \begin{minipage}{13cm}
    $^1$ Department of Mathematics and Statistics,\\
    The University of Melbourne, Parkville, Vic 3010, Australia.\\
    \texttt{owczarek@unimelb.edu.au}\\[1ex]
$^2$ School of Mathematical Sciences\\
Queen Mary University of London\\
Mile End Road, London E1 4NS, UK\\
\texttt{t.prellberg@qmul.ac.uk}
\end{minipage}
}

\maketitle

\begin{abstract}
  Self-avoiding walks and self-avoiding trails, two models of a
  polymer coil in dilute solution, have been shown to be governed by
  the same universality class. On the other hand, self-avoiding walks
  interacting via nearest-neighbour contacts (ISAW) and self-avoiding
  trails interacting via multiply-visited sites (ISAT) are two models
  of the coil-globule, or collapse transition of a polymer in dilute
  solution. On the square lattice it has been established numerically that
  the collapse transition of each model lies in a different
  universality class.

  The models differ in two substantial ways. They differ in the types
  of subsets of random walk configurations utilised (site self-avoidance
  versus bond self-avoidance) and in the type of attractive interaction.
  It is therefore of some interest to consider self-avoiding trails
  interacting via nearest neighbour attraction (INNSAT) in order to ascertain
  the source for the difference in the collapse universality class.
  Using the flatPERM algorithm, we have performed computer simulations of this
  model. We present numerical evidence that the singularity in the free energy
  of INNSAT at the collapse transition has a similar exponent to that of the
  ISAW model rather than the ISAT model. This would indicate that the type of
  interaction used in ISAW and ISAT is the source of the difference in universality
  class.

\end{abstract}
\newpage

\section{Introduction}
\label{sec:introduction}

The collapse transition of a polymer in a dilute solution has been a
focus of study in lattice statistical mechanics for decades
\cite{gennes1975a-a,gennes1979a-a}. Any lattice model of a collapsing
polymer has two key ingredients: an excluded volume effect expressing
the impenetrability of monomers, and a short-range attractive force,
which mimics the complex monomer-solvent interaction. When the effects of
the excluded volume and the short-range attraction balance each other, the
polymer undergoes a collapse transition which separates two distinct phases:
a swollen and a collapsed phase.

The canonical lattice model of the configurations of a polymer in
solution has been the model of self-avoiding walks (SAW) where a random walk
on a lattice is not allowed to visit a lattice site more than once. SAW
display the desired excluded volume effect and are swollen in size
relative to unrestricted random walks at the same length. A common way
to introduce a short-range interaction is to assign a negative energy
to each non-consecutive pair of monomers lying on neighbouring lattice
sites, modelling an effective attractive force. This is the
interacting self-avoiding walk (ISAW) model, which is the standard
lattice model of polymer collapse using self-avoiding walks.

The properties of lattice polymers are also related to those of magnetic
systems near their critical point \cite{gennes1972a-a}. More precisely,
lattice polymers are related to magnets with $O(n)$ symmetry in the formal limit
of zero components ($n \to 0$). This relation is
of great importance, since it allows the application of the methods of
statistical field theory to the study of polymer models, and
the collapse transition
can be understood as the tri-critical point of such systems \cite{gennes1975a-a,
stephen1975a-a,duplantier1982a-a}.

The study of the critical properties of lattice polymers, and thus of $O(n)$
models when we let $n \to 0$, in two dimensions has been ongoing over
decades theoretically and numerically. Nienhuis in 1982 \cite{nienhuis1982a-a}
was able to compute the critical exponents of free SAW by considering a model
of non-intersecting loops on the hexagonal lattice, and in 1987 Duplantier and
Saleur \cite{duplantier1987a-a}
were able to model the bond interaction introducing vacancies on the same
lattice, obtaining a full set of critical exponents  for the
polymer collapse transition in the ISAW model.
Their conjectured values for the exponents have been subsequently confirmed
numerically by Prellberg and Owczarek in 1994 \cite{prellberg1994a-:a}. We will
refer to (the universality class of) this critical point as the `$\theta$-point'.

Introducing attractive interactions between bonds opens the doors to more complex
possibilities. In the quest for a solvable $O(n)$ model on the square lattice,
Bl\"ote and Nienhuis in 1989 \cite{blote1989a-a} considered a lattice model,
which includes energies for site-collisions and straight segments. For this
model five critical branches are exactly known
\cite{blote1989a-a, Batchelor:1989ji,nienhuis1990a-a,warnaar1992b-a}. One of
these branches (named `branch 0' in \cite{blote1989a-a})
is similar to the $\theta$-point previously found by Duplantier and Saleur \cite{duplantier1987a-a} (for this
branch Batchelor \cite{batchelor1993a-a} obtained the exponents $\nu = 4/7$,
$\gamma = 6/7$), and two branches correspond to dense and dilute (SAW) polymers.
The two remaining branches are, respectively, associated with a combination
of Ising-like and $O(n)$ critical behaviour and with a new tri-critical point.
This new tri-critical point (with exponents $\nu = 12/23$, $\gamma = 53/46$)
is another candidate for describing a collapsing polymer.

A different model of a collapsing model can be constructed starting
from self-avoiding trails. A self-avoiding trail (SAT) is a lattice
walk configuration where the excluded volume is obtained by preventing
the walk from visiting the same bond, rather than the same site, more
than once. This is a slightly weaker restriction, and SAW
configurations are a proper subset of SAT configurations. The
interacting version of self-avoiding trails (ISAT), customarily
obtained by giving an energy to multiple visited sites, also presents
a collapse transition.

It is known that SAW and SAT share the same statistics in their
high-temperature phase, but theoretical prediction and numerical
evidence \cite{owczarek1995a-:a,owczarek2007c-:a} strongly suggests
that the collapse transition of the ISAT model is in a different
universality class to that of ISAW, although there is no clear
understanding of why this would be the case. An interesting possible
explanation of the ISAT collapse has been provided by Foster
\cite{foster2009a-a} in terms of the tri-critical point with exponents
$\nu = 12/23$, $\gamma = 53/46$ derived from the $O(n)$ model.

Although they both aim to describe the same physical system, the ISAW
and ISAT models differ in both their geometrical properties and their
interaction.  Their different critical behaviour could be due to
either one of these two differences.

To investigate this further, we considered a mixed model where trails
interact in the same way as SAW, that is, by a nearest-neighbour
interaction. We shall call this new hybrid model Interacting
Nearest-Neighbour Self-Avoiding Trails, or INNSAT. In the next section
we review the ISAW and ISAT models describing the different behaviours
that they have been shown to demonstrate. In Section~\ref{sec:model}
we formally introduce our new model INNSAT and the quantities of
interest. We then describe our results in Section~\ref{sec:results}
and summarise our conclusions in the final section.

\section{ISAW and ISAT}
\subsection{Interacting Self-Avoiding Walks (ISAW)}
\label{sec:isaw}

Let us recall briefly the definition and main properties of the ISAW
model.  Consider the ensemble $\mathcal S_n$ of self-avoiding walks
(SAW) of length $n$, that is, of all lattice paths of $n$ steps that
can be formed on the square lattice such that they never visit the
same site more than once.
Given a SAW $\varphi_n \in \mathcal S_n$, we define a \emph{contact}
whenever there is a pair of sites that are neighbours on the lattice
but not consecutive on the walk. We associate an energy
$-\varepsilon_c$ with each contact.
Denoting by $m_c(\varphi_n)$ the number of contacts in $\varphi_n$,
the probability of $\varphi_n$ is given by
\begin{equation}
	\frac{e^{\beta \varepsilon_c m_c(\varphi_n)}}{Z_n(T)},
\end{equation}
and the partition function  $Z_n(T)$ is defined in the usual way as
\begin{equation}
	Z_n(T) = \sum_{\varphi_n\in\mathcal S_n}\ e^{\beta \varepsilon_c m_c(\varphi_n)},
\end{equation}
where $\beta$ is the inverse temperature $1/k_B T$ ($k_B$ is
Boltzmann's constant).
We define a Boltzmann weight (fugacity) $\omega_c = \exp(\beta
\varepsilon_c)$.  The finite-length reduced free energy is
\begin{equation}
	\kappa_n(T) = \frac{1}{n} \log\ Z_n(T)
\end{equation}
and the thermodynamic limit is obtained by taking the limit of large $n$, i.e.,
\begin{equation}
	\kappa(T) = \lim_{n \to \infty} \kappa_n(T).
\end{equation}
As mentioned above, it is expected that there is a collapse phase
transition at a temperature $T_\theta$, which is known as the
$\theta$-point, characterised by a non-analyticity in $\kappa(T)$.

The temperature $T_\theta$ also separates regions of different
finite-length scaling behaviour for fixed temperatures. Considering
this finite-length scaling, for high temperatures ($T > T_\theta$) the
excluded volume interaction is the dominant effect, and the behaviour
is universally the same as for the non-interacting SAW problem: for
large $n$, the mean squared end-to-end distance (or equivalently the
radius of gyration) $R_n^2$ and partition function $Z_n$ are expected
to scale as
\begin{align}
	R_n^2 & \sim A \,n^{2 \nu} \quad \text{with $\nu > 1/2$}\quad\text{ and} \\
	Z_n   & \sim D\, \mu^n n^{\gamma-1},
\end{align}
respectively, where $\log \mu = \kappa(T)$ and the exponents $\nu$ and
$\gamma$ are expected to be universal. In two dimensions it is well
established \cite{nienhuis1982a-a} that $\nu=3/4$ and $\gamma=43/32$
for $T>T_\theta$. The constants $A$ and $D$ are temperature dependent.

Fixing the temperature precisely at the $\theta$-point, $T=T_\theta$,
Duplantier and Saleur \cite{duplantier1987a-a} found $\nu = 4/7$ and
$\gamma = 8/7$ in two dimensions.

For low temperatures ($T < T_\theta$) it is accepted that the
partition function is dominated by configurations that are internally
dense, though not necessarily fully dense. The partition function
should then scale differently from that at high temperatures, since a
collapsed polymer should have a well-defined surface (and associated
surface free energy) \cite{owczarek1993c-:a}. One expects in $d$
dimensions large-$n$ asymptotics of the form
\begin{align}
	R_n^2 & \sim A \,n^{2/d}\quad\text{and} \\
	Z_n   & \sim D \,\mu^{n} \mu_s^{n^{(d-1)/d}} n^{\gamma'-1},
\end{align}
with $\mu_s < 1$. The constants $A$ and $D$ are temperature
dependent. It is expected that the internal density smoothly goes to
zero as the temperature is raised to the $\theta$-point.

To explore the singularity in the free energy at the collapse point
further, it is useful to consider the (reduced) internal energy and
the specific heat, which are defined as
\begin{equation}
	u_n(T) = \frac{\< m_c \>}{n} \quad \mbox{ and } \quad
	c_n(T) = \frac{\<m_c^2\> - \<m_c\>^2 }{n}
	,
\end{equation}
with limits
\begin{align}
	U(T) = \lim_{n \to \infty} u_n(T) \quad \mbox{ and } \quad
	C(T) = \lim_{n \to \infty} c_n(T)
	.
\end{align}
When $T\to T_\theta$, the singular part of the specific heat behaves as
\begin{equation}
	C(T) \sim B\, \left|T_\theta - T\right|^{-\alpha} ,
\end{equation}
where $\alpha < 1$ for a second-order phase transition. If the
transition is second-order, the singular part of the thermodynamic
limit internal energy behaves as
\begin{equation}
	U(T) \sim B\, \left|T_\theta - T\right|^{1-\alpha}
\end{equation}
when $T\to T_\theta$, and there is a jump in the internal energy at
$T_\theta$ if the transition is first-order (an effective value of
$\alpha = 1$).


Tri-critical scaling \cite{brak1993a-:a} predicts that around the
critical temperature, the finite-length scaling of the singular part
of the specific-heat $c_n(T)$ obeys the following crossover scaling
form
\begin{equation}
	c_n(T) \sim n^{\alpha \phi}\ \mathcal C\big( (T-T_\theta) n^{\phi} \big) ,
\end{equation}
when $T\to T_\theta$ and $n\to\infty$, and that the exponents $\alpha$
and $\phi$ are related via
\begin{equation}
	2 - \alpha = \frac{1}{\phi}.
\end{equation}
If one considers the peak of the finite-length specific heat it will
behave as
\begin{equation}
	c_n^{peak} \sim n^{\alpha \phi}\ \mathcal C\big( x^{max}),
\end{equation}
where $x^{max}$ is the location of the maximum of the function
$\mathcal{C}(x)$.

The work of Duplantier and Saleur (1987) predicts the exponents for
the $\theta$-point collapse are
\begin{equation}
  \phi = \phi_{\theta} = 3/7
\end{equation}
and
\begin{equation}
  \alpha = \alpha_{\theta}= - 1/3 .
\end{equation}
It is important to observe that this implies that the specific heat
does not diverge at the transition since the exponent
\begin{equation}
  \alpha_{\theta} \phi_{\theta} =-1/7\approx - 0.14
\end{equation}
is negative. However, the peak values of the third derivative of the
free energy with respect to temperature will diverge with positive
exponent
\begin{equation}
  (1 + \alpha_{\theta}) \phi_{\theta} = 2/7\approx0.28.
\end{equation}

\subsection{Interacting Self-Avoiding Trails (ISAT)}
\label{sec:isat}

The model of interacting trails on the square lattice is defined as
follows. Consider the ensemble $\mathcal T_n$ of self-avoiding trails
(SAT) of length $n$, that is, of all lattice paths of $n$ steps that
can be formed on the square lattice such that they never visit the
same bond more than once.
Given a SAT $\psi_n \in \mathcal T_n$, we associate an energy
$-\varepsilon_t$ with each doubly visited site.  Denoting by
$m_t(\psi_n)$ the number of doubly visited sites in $\psi_n$, the
probability of $\psi_n$ is given by
\begin{equation}
	\frac{e^{\beta \varepsilon_t m_t(\psi_n)}}{Z^{ISAT}_n(T)},
\end{equation}
where we define the Boltzmann weight $\omega_t = \exp(\beta
\varepsilon_t)$ and the partition function of the ISAT model is given
by
\begin{equation}
	Z^{ISAT}_n(T) = \sum_{\psi_n\in \mathcal T_n}\ \omega_t^{m_t(\psi_n)}.
\end{equation}
Previous work \cite{owczarek1995a-:a,owczarek2007c-:a} on the square
lattice has shown that there is a collapse transition at a temperature
$T=T_t$ with a strongly divergent specific heat, and the exponents
have been estimated as
\begin{equation}
  \phi_{\text{\sc it}} =0.84(3)\quad \mbox{ and } \quad \alpha_{\text{\sc it}}=0.81(3)\;,
\end{equation}
arising from a scaling of the peak value of the specific heat
diverging with exponent
\begin{equation}
  \phi_{\text{\sc it}}  \alpha_{\text{\sc it}}=0.68(5)\;.
\end{equation}
This result is a clear difference to the ISAW $\theta$-point described
above where the singularity in the specific heat is \emph{convergent}.
Additionally, at $T=T_t$ the finite-length scaling of the end-to-end
distance was found to be consistent \cite{owczarek1995a-:a} with the
form
\begin{equation}
  R_n^2(T) \sim A n\left(\ln n\right)^2
\end{equation}
as $n\to\infty$. Again, this is quite different to the exponent
$\nu=4/7$ for the ISAW.

Another important difference is that it has been recently observed
\cite{doukas2010a-:a,bedini2012=a-:a} that the low temperature phase
is maximally dense on the triangular and square lattices. On the square lattice this implies that if one
considers the proportion of the sites on the trail that are at
lattice sites which are not doubly occupied via
\begin{equation}
  p_n=\frac{n - 2 \< m_t \>}{n},
\end{equation}
then it is expected that
\begin{equation}
  p_n \rightarrow 0 \quad \mbox{as} \quad n\rightarrow \infty.
\end{equation}

\section{The INNSAT model}
\label{sec:model}

We define our new interacting model of trails (INNSAT) as
follows. Consider the set of bond-avoiding paths $\mathcal T_n$ as
defined in the previous section. When two sites are adjacent on the
lattice but not consecutive along the walk, so as not to be joined by
{\it any} step of the walk, we again refer to this pair of sites as a
nearest-neighbour \emph{contact} and we give it a weight
$\omega_c=e^{\beta \varepsilon_c}$, analogously to the ISAW model. In
Figure~\ref{fig:nnsat} a trail of length $n=21$ with $m_c=6$ contacts
is illustrated.

Denoting by $m_c(\psi_n)$ the number of contacts in $\psi_n$, the
probability of $\psi_n$ is given by
\begin{equation}
	\frac{e^{\beta \varepsilon_c m_c(\psi_n)}}{Z^{\text{\sc nt}}_n(T)},
\end{equation}
where the partition function is
\begin{equation}
  Z^{\text{\sc nt}}_n(T) = \sum_{\psi_n\in\mathcal T_n}\ \omega_c^{m_c(\psi_n)}.
\end{equation}
The intensive reduced internal energy  and specific heat are as for ISAW:
\begin{equation}
	u_n(T) = \frac{\< m_c \>}{n} \quad \mbox{ and } \quad
	c_n(T) = \frac{\<m_c^2\> - \<m_c\>^2 }{n} .
\end{equation}
We shall also consider the
proportion of the sites on the trail that are at lattice sites which are not doubly occupied via
\begin{equation}
p_n=\frac{n - 2 \< m_t \>}{n},
\end{equation}
where $m_t$ is the number of doubly visited sites as defined above.

We shall see there is a collapse transition at a single temperature, $T_p$, equivalently, fugacity $\omega^p_c$.  If the INNSAT model behaves in the same way as the ISAW model we would not expect the specific heat to diverge at this collapse point so we also need to introduce a quantity $t_n$ proportional to the third derivative of the free energy as
\begin{equation}
	t_n = \frac{\<m_c^3\> - 3\<m_c\>\<m_c^2\> + 2\<m_c\>^3}{n}
	.
\end{equation}
This quantity should have a singular part that behaves as
\begin{equation}
	t_n(T) \sim n^{(\alpha +1)\phi}\ \mathcal T\big( (T-T_p) n^{\phi} \big) ,
\end{equation}
around the collapse point, i.e., as $T\to T_p$ and $n\to\infty$. Hence the peaks of this quantity $t_n$ will scale with exponent $(\alpha +1)\phi$.

\begin{figure}
  \begin{center}
  \begin{tikzpicture}
    \draw[help lines] (-1,-3) grid (5,4); \fill (0,0) circle (2pt);
    \draw[red,thick,decorate,decoration={zigzag,segment length=2mm}] (1,-1)
    -- (2,-1) (2,1) -- (1,1) (2,2) -- (2,3) (3,2) -- (3,3) (4,2) --
    (4,3) (1,2) -- (2,2);
    \draw[->,very thick,rounded corners=5pt] (0,0) -- ++(1,0) -- ++(0,-2)
    -- ++(1,0) -- ++(0,4) -- ++(2,0) -- ++(0,-2) -- ++(-3,0) -- ++(0,3)
    -- ++(3,0);
  \end{tikzpicture}
  \end{center}
  \caption{An example of INNSAT configuration with $m_c = 6$, that is, there are $6$ nearest-neighbour contacts illustrated via zigzag (red) lines. The trail can visit a site of the lattice twice by ``touching" and by ``crossing" itself. The number of doubly visited sites is $m_t=2$. Note that there is no contact between the second and the seventh visited site of the walk, even though these are non-consecutive nearest-neighbour sites, as both sites are visited consecutively by a different segment of the trail. }
\label{fig:nnsat}
\end{figure}

\section{Numerical Results}
\label{sec:results}

We have first simulated the INNSAT model using the flatPERM algorithm
\cite{prellberg2004a-a} up to length $n = 1000$. With $S_n \simeq 2.7
\cdot 10^6$ iterations, we collected $2.6 \cdot 10^{10}$ samples at
the maximum length. Following \cite{prellberg2004a-a}, we also
measured the number of samples adjusted by the number of their
independent growth steps, obtaining $S_n^{eff} \simeq 1.6\cdot 10^8$
``effective samples''.

FlatPERM outputs an estimate $W_{n,m_c}$ of the total weight of the
walks of length $n$ at fixed values of $m_c$. From the total weight
one can access physical quantities over a broad range of temperatures
through a simple weighted average, e.g.
\begin{align}
	\< \mathcal O \>_n(\omega) = \frac{\sum_{m_c} \mathcal O_{n,m_c}\, \omega^{m_c} \, W_{n,m_c}}{\sum_m \omega^{m_c} \, W_{n,m_c}}.
\end{align}


We have begun by analysing the scaling of the specific heat by
calculating the location of its peak $\omega_n^p =
\argmax_{\omega}\ c_n(\omega)$ and thereby evaluating $c_n^p =
c_n(\omega^p_n)$. We have also found the two peaks of $t_n(T)$ as
a function of temperature: the peak values we denote $t^{p,\pm}_n$.

\begin{figure}[ht!]
  \begin{center}
    \includegraphics[width=0.45\columnwidth]{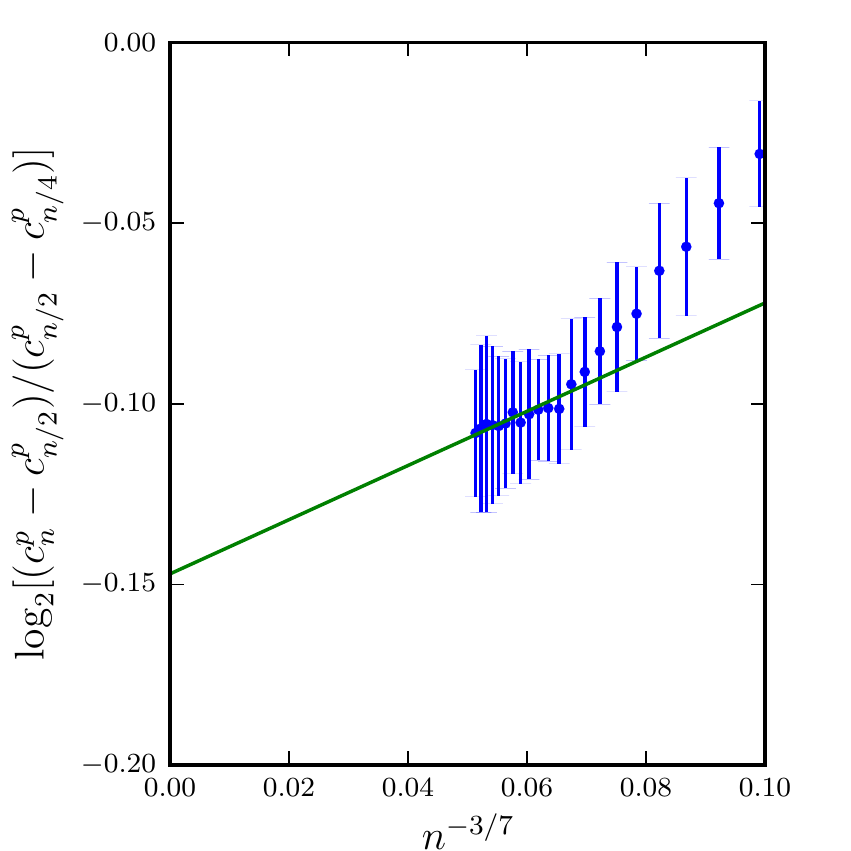}
    \includegraphics[width=0.45\columnwidth]{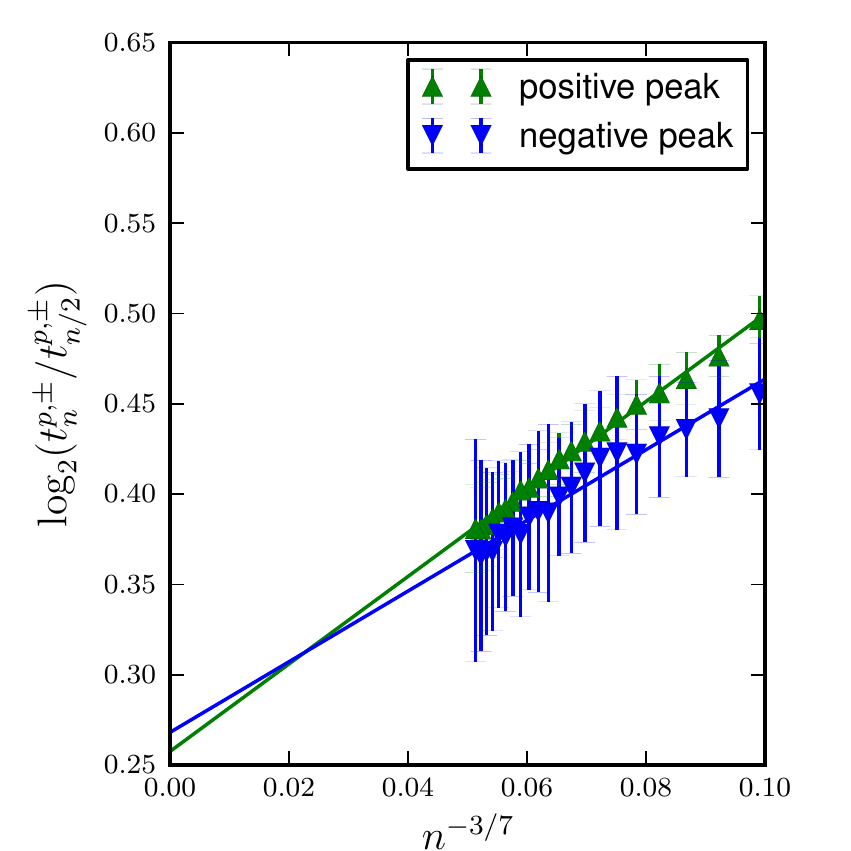}
  \end{center}
  \caption{Left: A plot of (\ref{quantity}) designed to estimate the
    exponent $\alpha\phi$ from the specific heat. Right: A plot of
    (\ref{quantity2}) designed to estimate the exponent
    $(1+\alpha)\phi$ from the derivative of the specific heat. The
    upper (green) points are values from the negative peak of $t_n$
    while the lower (blue) points are from the positive peak of
    $t_n$.}
  \label{fig:sp_and_tr}
\end{figure}

It is clear that the specific heat peak is growing rather weakly as
length increases. As the specific heat might converge, we obtain
finite size estimates of the specific heat exponent by considering
\begin{equation}
  \label{quantity}
  \log_2\left[\frac{c^p_n- c^p_{n/2}}{c^p_{n/2}-c^p_{n/4}}\right]
\end{equation}
which should converge to $\alpha\phi$ as $n\to\infty$. Assuming
corrections to scaling of $n^{-3/7}$ as for ISAW, we find
\begin{equation}
\alpha_{\text{\sc nt}}\phi_{\text{\sc nt}} = -0.15(5),
\end{equation}
as shown on the left-hand side of Figure~\ref{fig:sp_and_tr}.  This is
consistent with a value of $-1/7\approx-0.14$ for the $\theta$-point
universality class.

Furthermore when we analyse $t_n$, which is proportional to the third
derivative of the free energy, we find further consistent values of
exponents. This derivative has two peaks $t^{p,\pm}_n$, and we obtain
finite size estimates of the specific heat exponent from each of the
peaks by considering
\begin{equation}
  \label{quantity2}
  \log_2\left[\frac{t^{p,\pm}_n}{t^{p,\pm}_{n/2}}\right]
\end{equation}
which should converge to $(1+\alpha)\phi$ as $n\to\infty$. From the
plot on the right-hand side of Figure~\ref{fig:sp_and_tr} we estimate
\begin{equation}
  (\alpha_{\text{\sc nt}}+1) \phi_{\text{\sc nt}} = 0.26(5).
\end{equation}
There is still considerable curvature in the right-hand plot which
would indicate a slightly larger value when higher order corrections
to scaling are taken in account. Once again we conclude that the
$\theta$ value of $2/7\approx 0.28$ is consistent with our
data.

We therefore conjecture that
\begin{equation}
  \phi_{\text{\sc nt}} = \phi_{\theta}\neq \phi_{\text{\sc it}} \quad \mbox{ and } \quad \alpha_{\text{\sc nt}}=\alpha_{\theta}\neq\alpha_{\text{\sc it}}\;.
\end{equation}

%
\begin{figure}[ht!]
\begin{center}
\includegraphics[width=0.45\columnwidth]{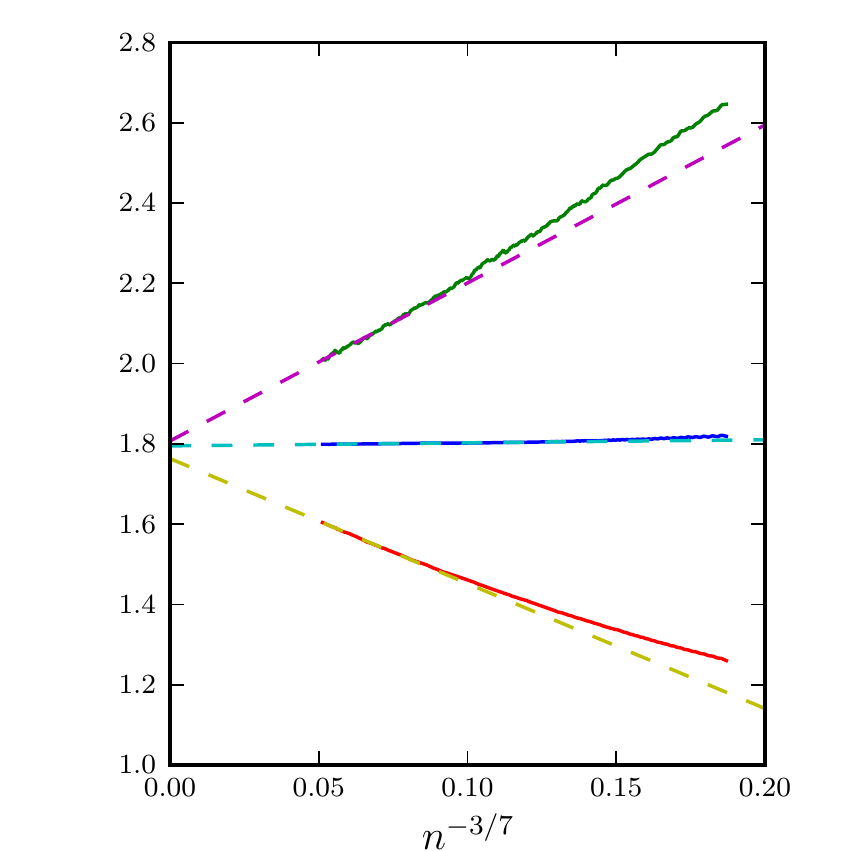}
\end{center}
\caption{A plot of the locations of the two peaks of $t_n(T)$ and of
  $c_n(T)$ against $n^{-3/7}$.  Extrapolation for $n\to\infty$
  provides an estimate for the thermodynamic transition temperature
  $\omega^p$. The upper (green) curve is the location of negative
  $t^{p,-}_n$ peak, the lower (red) curve is the location of positive
  $t^{p,+}_n$ peak while the middle (blue) curve is the location of
  the peak $c^p_n$ of the specific heat.}
\label{fig:tclocate}
\end{figure}
Using the locations of the peaks of the specific heat and its derivative and assuming finite-size correction of the order of $n^{-3/7}$, we have estimated the location of the collapse point as indicated in Figure~\ref{fig:tclocate}. The magnitude of the difference between the three peak positions at finite length indicates the large corrections to scaling present in this problem at length $1000$. We estimate
\begin{equation}
\omega^{p} = 1.78(1).
\end{equation}

We next analysed the scaling of the end-to-end distance of the polymer.
To obtain an estimate for the exponent $\nu$ we considered a finite-size effective exponent
\begin{equation}
\label{quantity3}
	\nu^{est}_n(\omega) = \frac{1}{2} \log_2\left[ \frac{\<R^2\>_n(\omega)}{\<R^2\>_{n/2}(\omega)}\right]
	.
\end{equation}

\begin{figure}[ht!]
  \begin{center}
    \includegraphics[width=0.45\columnwidth]{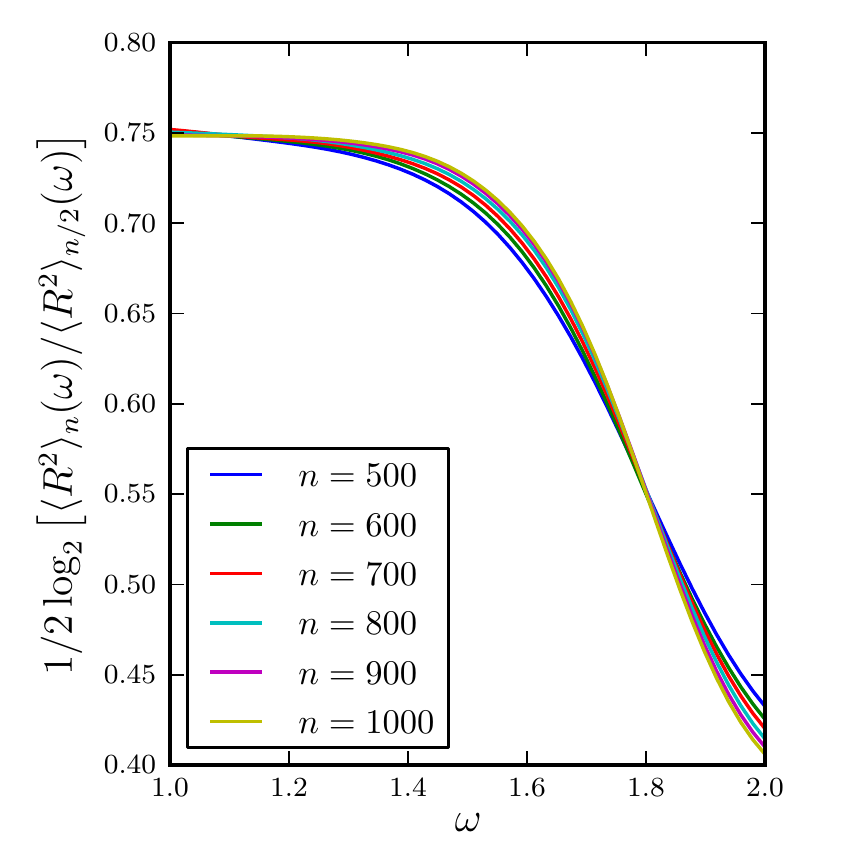}
    \includegraphics[width=0.45\columnwidth]{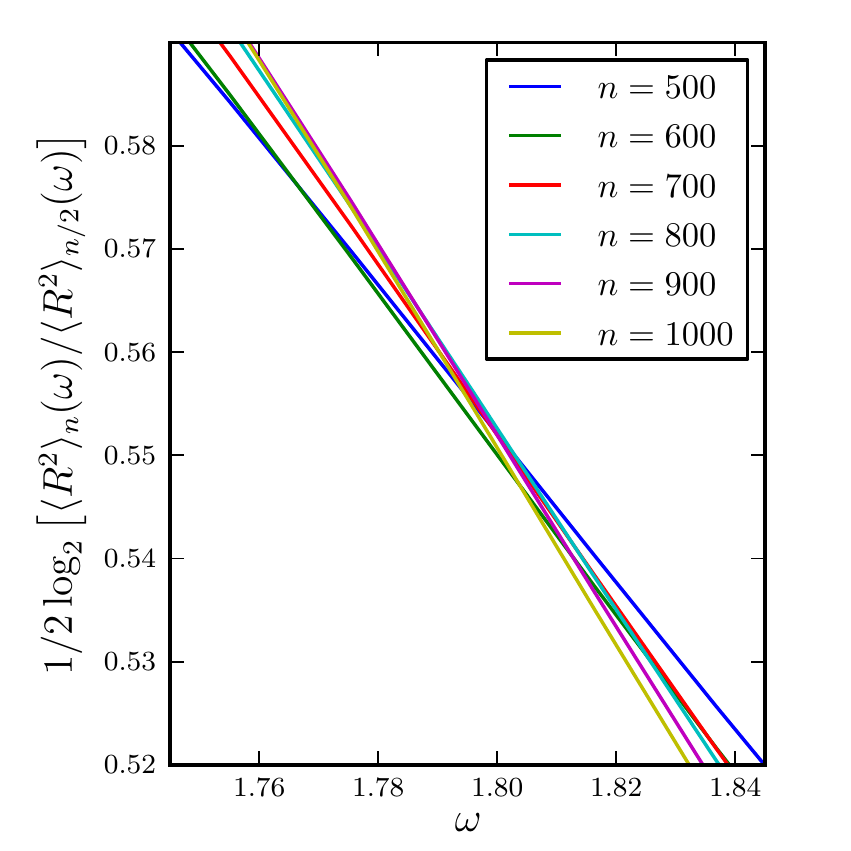}
  \end{center}
  \caption{Estimates of the exponent $\nu$ for INNSAT from finite-size
    estimates (\ref{quantity3}) as a function of temperature (left)
    and a zoom (right) on the crossing region. There are crossing points in the region of $\omega$ between $1.78$ and $1.80$ giving corresponding estimates of $\nu$ between $0.57$ and $0.55$.
  }
  \label{fig:radius-nu}
\end{figure}

\begin{figure}[ht!]
  \begin{center}
    \includegraphics[width=0.45\columnwidth]{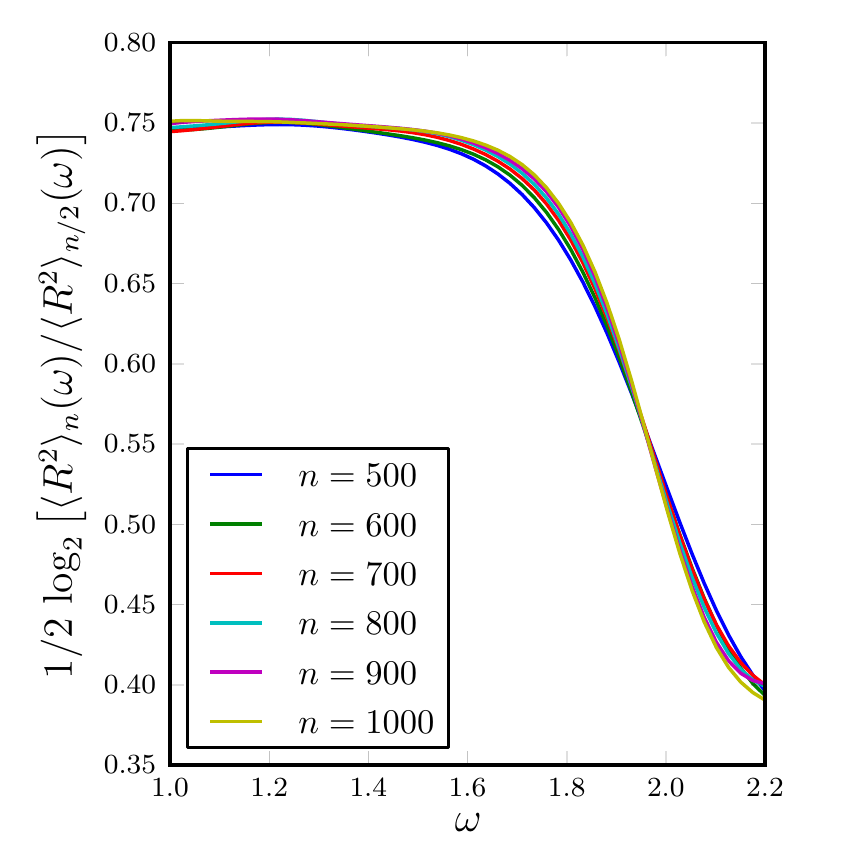}
    \includegraphics[width=0.45\columnwidth]{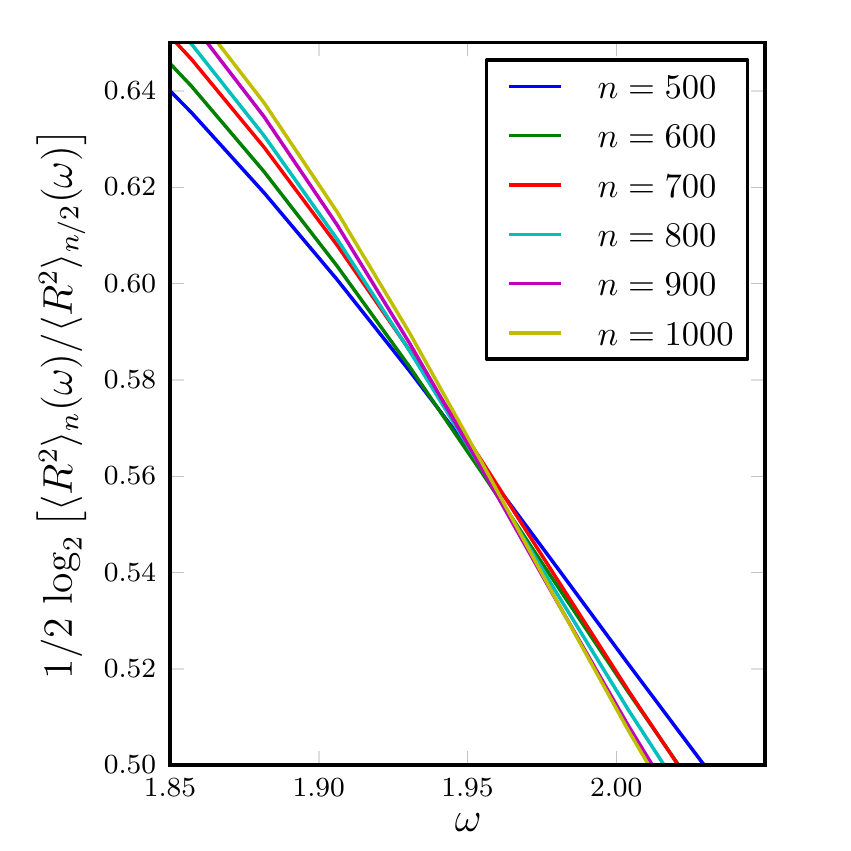}
  \end{center}
  \caption{Estimates of the exponent $\nu$ for ISAW from finite-size
    estimates (\ref{quantity3}) as a function of temperature (left)
    and a zoom (right) on the crossing region. The crossing point is
    at $\omega \approx 1.96$ and $\nu \approx0.56$. While the estimate
    for $\nu$ at the crossing point is closer to the value of $4/7$
    than the one obtained for INNSAT, the difference provides evidence
    for the presence of fairly strong corrections-to-scaling.}
  \label{fig:radius-nu-saw}
\end{figure}

Using the data obtained from our flatPERM simulations, we plotted this
quantity against temperature for various values of the length $n$, as
shown in Figure~\ref{fig:radius-nu}. We find that the graphs for
different values of $n$ intersect around a particular value of the
temperature. The location of this intersection point is a good
estimator of the infinite-length critical temperature $\omega^p$ and of
the exponent $\nu$ at the transition. We find a region of crossing points in $\omega$ between $1.78$ and $1.80$ with corresponding estimates of  $\nu$ between $0.57$ and $0.55$. While the obtained exponent estimate range is a little below the $\theta$ point value of $4/7\approx0.571$, we
expect that our estimate is affected by strong corrections to
scaling. For comparison, we simulated the ISAW model and performed an
identical analysis, shown in Figure~\ref{fig:radius-nu-saw}. The ISAW
estimate of $\nu \approx 0.56$ at the collapse transition, while being
closer to 4/7, also indicates the presence of strong corrections to
scaling.

\begin{figure}[ht!]
  \begin{center}
    \includegraphics[width=0.6\columnwidth]{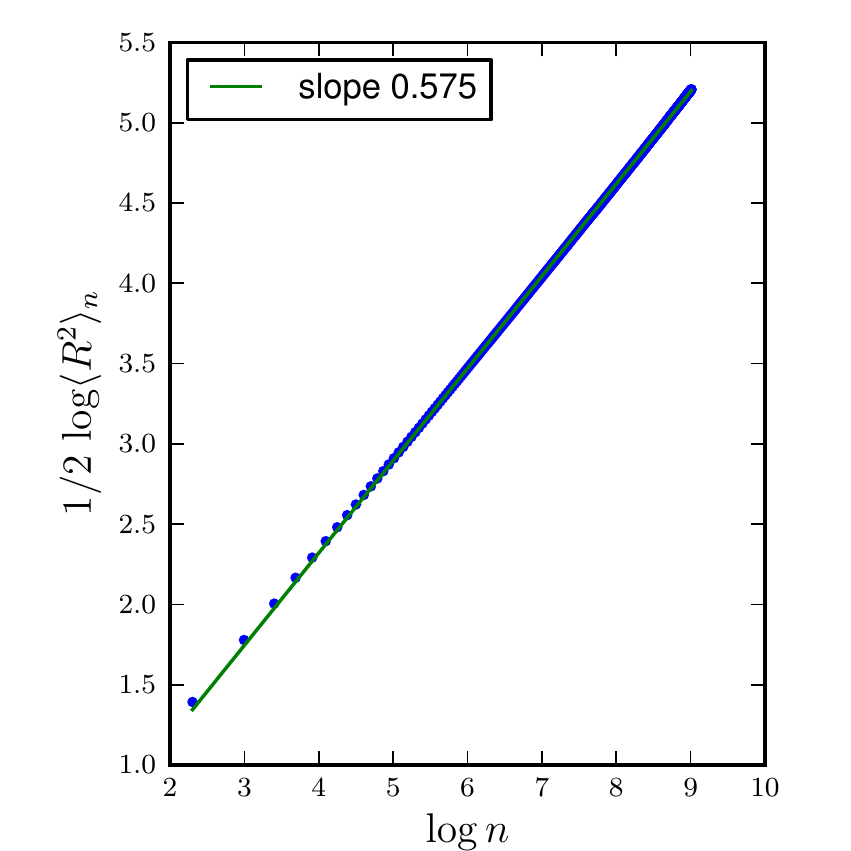}
  \end{center}
  \caption{Estimate of the exponent $\nu$ from from finite-size
    estimates (\ref{quantity3}) obtained from a thermal simulation at
    $\omega = 1.78$ up to length $n = 8192$. This estimate is
    sensitive to the estimate of the critical fugacity. A lower
    critical estimate would lead to a higher exponent estimate.}
  \label{fig:radius-thermal}
\end{figure}

Hence, we ran a thermal simulation at the estimated critical fugacity
of $1.78$ up to $n = 8192$. With $S = 5 \cdot 10^6$ iterations, we
obtained $S_n \simeq 1.8 \cdot 10^7$ samples at the maximum length
(corresponding to $S_n^{eff} \simeq 1.2 \cdot 10^5$ ``effective
samples''). Figure~\ref{fig:radius-thermal} shows a log-log plot of
the radius with respect to the size of the walk along with a linear
best fit with slope 0.575. Of course, using $\omega=1.77$ or
$\omega=1.79$ gives estimates of $\nu$ that differ by $0.01$ so the
sensitivity to the location of the critical point dominates the error
of our estimate.
Correspondingly, if we assume that the INNSAT is in the ISAW
universality class we can use the value of $\nu=4/7$ to better
estimate the critical fugacity in the INNSAT model as
\begin{equation}
  \omega^{p} = 1.783(5).
\end{equation}

\subsection{Low-temperature phase}

\begin{figure}[ht!]
	\begin{center}
	\includegraphics[width=0.6\linewidth]{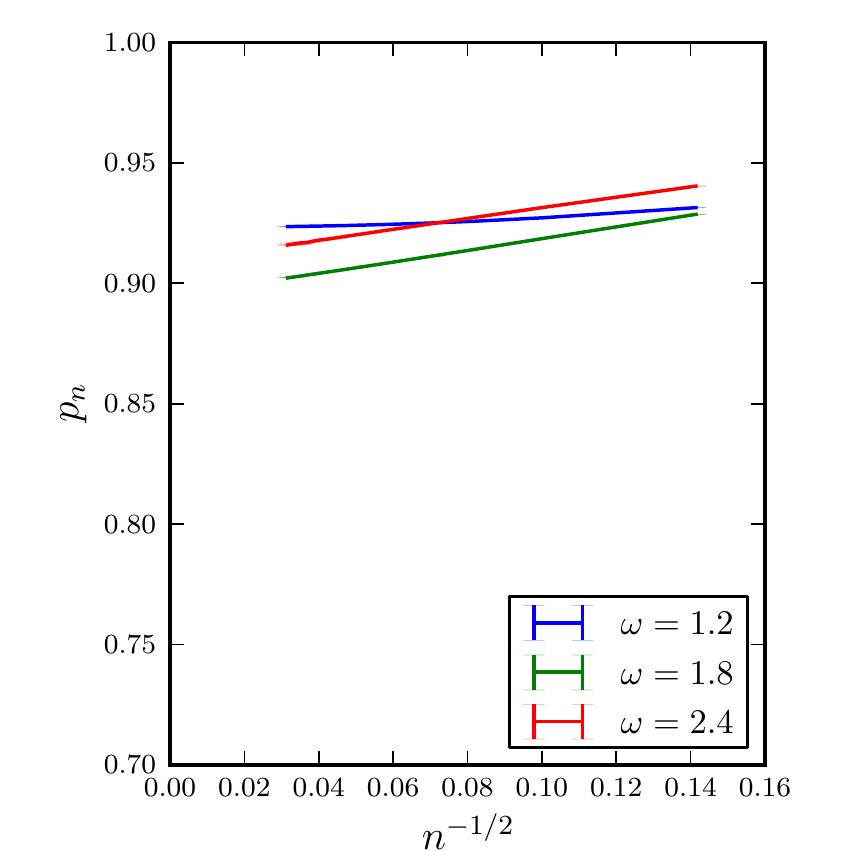}
	\end{center}
	\label{fig:low-temperature}
	\caption{Plots of $p_n$, the proportion of steps visiting the
          same site once, at a high temperature, upper (blue) curve,
          and a low temperature, lower (green) curve against
          $n^{-1/2}$. The scale $n^{-1/2}$ chosen is the natural low
          temperature scale. In both cases the asymptotic
          thermodynamic value is well away from zero. Errors are shown but are not visible at this scale.}
\end{figure}

There is strong evidence that the low-temperature phase of the ISAW
model is a globular phase that is not fully dense, while for
interacting trails the low-temperature phase is maximally dense. As
discussed above, the trail fills the lattice asymptotically in a
maximally dense phase, and the portion of steps not involved with
doubly-visited sites should tend to zero as $n \to \infty$. Following
the analysis in \cite{doukas2010a-:a} we measured the proportion $p_n$
of steps visiting the same site twice, and we plotted $p_n$ against
$n^{-1/2}$ at two different temperatures respectively above and below
the critical temperature, as shown in
Figure~\ref{fig:low-temperature}). We find that in both
low-temperature and high-temperature phases only a small portion of
the visited sites is visited again. The quantity $p_n$ stays close to
1 and, in particular, does not tend to zero in either region. This
phenomenon can be understood by noting that the nearest-neighbour
interaction makes energetically favourable for the walk to bounce away
from an already visited site rather than to visit it again.
This result implies that the low temperature phase is not fully dense,
just as for the ISAW low temperature phase. This provides a consistent
picture that the collapse transition in the INNSAT model is between
high and low temperature phases similar to the ISAW, making the
conclusion that the thermodynamic transition between them is similar
to the $\theta$-point a natural one.

\section{Conclusions}
\label{sec:conclusions}

We have considered a new lattice model of polymer collapse in two
dimensions which uses the interaction type, nearest neighbour
contacts, of the canonical lattice model (ISAW) but the configuration
space of an alternative model (ISAT), being bond-avoiding walks, also
known as self-avoiding trails. We find that the critical behaviour of
the free energy, and its derivatives of our model (INNSAT) seems to
align with the $\theta$-point universality class of the ISAW model
rather than that of the ISAT model, which has rather a strongly divergent 
specific heat not seen in either the ISAW or INNSAT models. The
end-to-end distance scaling is broadly consistent with this
conclusion, though estimates tend to be a little low. However, using
the assumption that the INNSAT and ISAW models lie in the same
universality class leads to a precise estimate of the critical
fugacity as $1.783(5)$. Given the different universality classes for
ISAT and INNSAT it may also be of interest to consider a model based
upon self-avoiding trail configurations with both types of interaction.

\section*{Acknowledgements}

Financial support from the Australian Research Council via its support
for the Centre of Excellence for Mathematics and Statistics of Complex
Systems and through its Discovery Program is gratefully acknowledged
by the authors. A L Owczarek thanks the School of Mathematical
Sciences, Queen Mary, University of London for hospitality.


\end{document}